\documentclass[aps,twocolumn,showpacs,superscriptaddress,amsmath,amssymb]{
revtex4-1}
\usepackage{graphicx}
\usepackage{color}
\usepackage{fancyhdr}

\newcommand{\eqreff}[1]{equation~(\ref{#1})}

\fancypagestyle{firststyle}
{ 

\setlength{\headsep}{1cm}
\fancyhf{}
\fancyhead[l]{Sci. Rep. 5, 7832; DOI:10.1038/srep07832 (2015) }
}

\begin{document}

\title{Local non-equilibrium thermodynamics}

\author{Lee Jinwoo}
\email{e-mail: jinwoolee@kw.ac.kr}
\affiliation{Department of Mathematics, Kwangwoon 
University, 20 Kwangwoon-ro, Nowon-gu, Seoul 139-701, Korea}

\author{Hajime Tanaka}
\email{e-mail: tanaka@iis.u-tokyo.ac.jp}
\affiliation{
Institute of Industrial Science, University of Tokyo, 4-6-1 Komaba, Meguro-ku, Tokyo 153-8505, Japan}

\date{Received \today}
\begin{abstract}
Local Shannon entropy lies at the heart of modern thermodynamics, with much discussion of trajectory-dependent entropy production. When taken at both boundaries of a process in phase space, it reproduces the second law of thermodynamics over a finite time interval for small scale systems. However, given that entropy is an ensemble property, it has never been clear how one can assign such a quantity \emph{locally}. Given such a fundamental omission in our knowledge, we construct a new ensemble composed of trajectories reaching an individual microstate, and show that locally defined entropy, information, and free energy are properties of the ensemble, or trajectory-independent true thermodynamic potentials. We find that the Boltzmann-Gibbs distribution and Landauer's principle can be generalized naturally as properties of the ensemble, and that trajectory-free state functions of the ensemble govern the exact mechanism of non-equilibrium relaxation.
\end{abstract}

\maketitle

\thispagestyle{firststyle}

\section*{Introduction}
\begin{figure*}
\centering
\includegraphics[width=10.5cm]{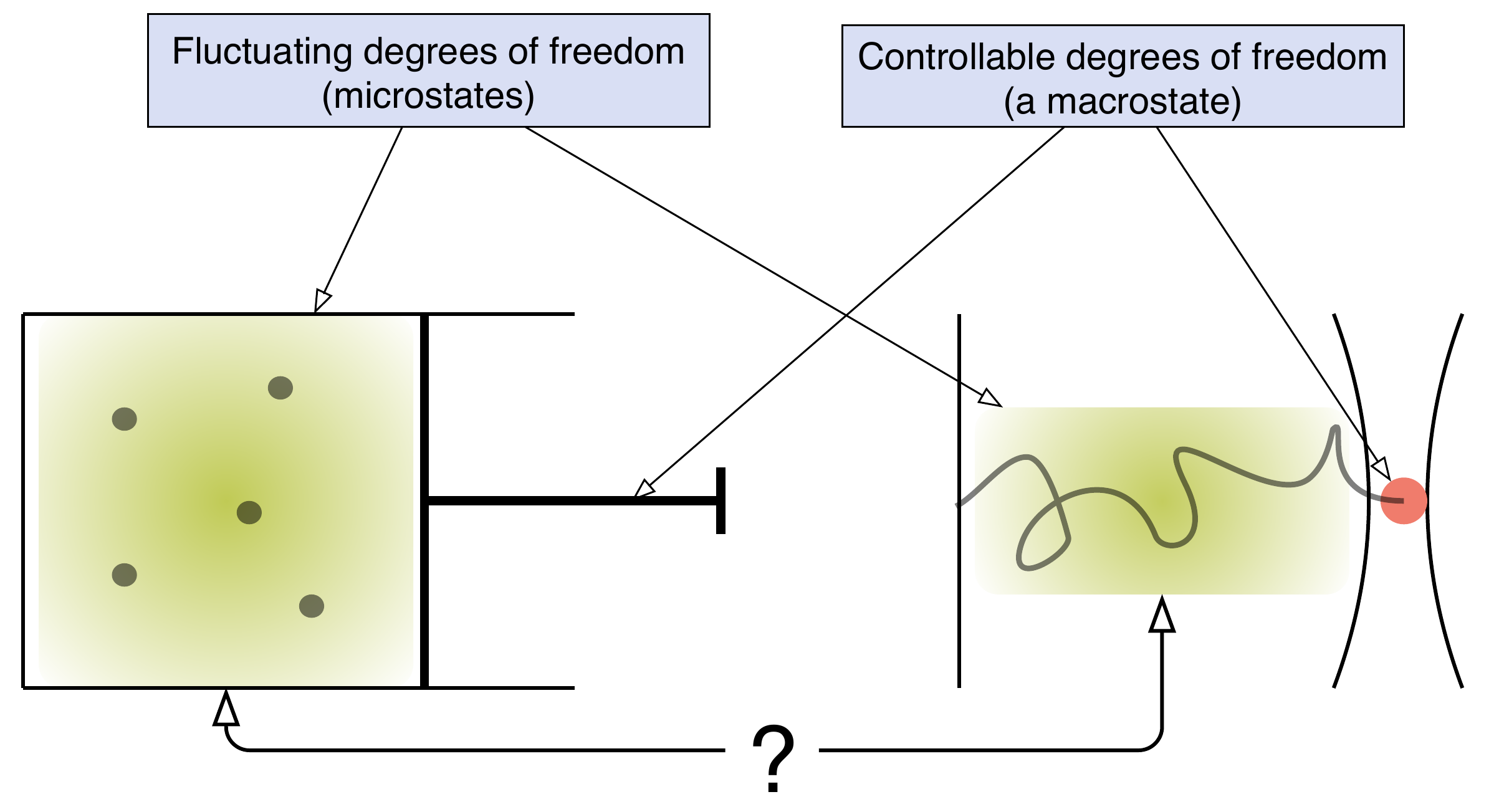}
\caption{{\bf A universal frame of thermodynamics and a fundamental question. } 
Thermodynamics is a theory for ensembles that are composed of a large number of fluctuating
degrees of freedom. An ensemble is constrained by a macrostate $\lambda$. 
In equilibrium thermodynamics, $\lambda$ varies
in a quasi-static manner. In the modern context, the theory considers arbitrary time-varying processes. 
Two typical systems in both approaches are shown schematically. On the left, gas particles
are fluctuating in a cylinder. The macrostate may be specified by the temperature, the volume of
the cylinder, and the number of particles. On the right, a polymer chain is fluctuating {under the influence of 
optically controlled bead connected to the end of the chain.} The macrostate may be specified by
the temperature, and the location of the bead {applying force to the molecule.
A fundamental question that we investigate is how we can justify 
thermodynamic descriptions of fluctuating microstates.}  
}
 \label{fig:fig1}
\end{figure*}

Statistical mechanics provides physical interpretations
of entropy and free energy 
that are macro-state functions 
({\it i.e.}, functions defined on a domain of the phase-space
points of a system, and thus inevitably non-local in character), 
 and sets bounds on permissible processes 
expressed as path functions like heat and work, 
{where a path is defined as a trajectory
of macrostates of a system \cite{callen}.}
As a system gets smaller, the effect of fluctuations
becomes significant, yet modern theory 
\cite{review,revSears,revSeifert} {provides permissible distributions 
of fluctuating path functions in the} form of beautiful
equalities \cite{morris1993,gala,kur,crooks99,sasa,kawai}.
{Modern theory has identified energetics \cite{sekimoto98} and
entropy production on the level of individual trajectories
\cite{seifert05}, and has linked path functions to properties of macrostates \cite{jar},
as verified experimentally \cite{hummer,liph2001,liph2002,expSasa,expColin,ritort2012}.
The relationship between classical and modern approaches} is schematically drawn in Fig.~\ref{fig:fig1}.

Despite these successes in 
non-equilibrium theories, however, still lacking is a unified framework 
analogous to classical thermodynamics \cite{jarReview}. 
{It is important to note that
contrary to the classical cases, a path 
in the modern theory
is defined as a trajectory of phase-space points of 
a system.
In general, the 
modern theory has been built effectively 
upon quantities defined
locally ({\it i.e.}, functions defined 
at each phase-space point). 
For example, a local form of information theoretic Shannon entropy, 
so called stochastic entropy, is defined as 
$-\ln p(x,t)$, where $p(x,t)$ is the phase-space density of a microstate $x$ of a system.  
Such a local form is necessary for boundary conditions of a process (a path) to recover
the second law of thermodynamics over a finite time interval for small scale systems \cite{crooks99,seifert05}. 
Then we immediately encounter a serious conceptual difficulty. As a component of the second law, the entropy should be a property of an ensemble. 
However, it has not been clear at all how such an ensemble property as entropy could be assigned \emph{locally}, or independently of a trajectory.  
This difficulty in the notion of locality is also closely associated with a classical view of difficulties of
non-equilibrium thermodynamics; the ill-defined notion of a macrostate during dynamic evolution 
of a system. It is thus difficult} to apply the language of equilibrium thermodynamics,
{most importantly the argument of counting}, to non-equilibrium problems.

Without counting, 
for both equilibrium and non-equilibrium {situations}, 
we have no way to even imagine what entropy is.
We remove this fundamental difficulty by 
constructing a new 
ensemble that is \emph{local} in space and time, and is well-defined even
in dynamic evolution of a system. 
Specifically, we count the accessible number of paths to a
microstate and consider all trajectories to a microstate in non-equilibrium
as an ensemble of the state.  
We, then,  show that local entropy, {similarly defined 
local information and local free energy are properties
of the ensemble, and relate  functions within
the ensemble to state functions of the ensemble}
which have been concealed behind the integrated-out forms of previous theories.

\section*{Theoretical Framework}
{In this article, we consider} a classical stochastic system in contact with a heat bath of temperature $T$. 
To provide {some intuitive grasp}, we take as a simple example a polymer chain whose one end is fixed and the other end under an external control $\lambda_{\tau}$
that may vary over time $0\le\tau\le t$ (see the right panel of Fig. 1).
{We will refer to $\lambda_{\tau}$ as a macrostate (the state of a polymer chain) at time $\tau$, 
and to the phase-space points of the polymer chain (excluding momentum variables for simplicity) as microstates. 
Note that, in this experiment, $\lambda_{\tau}$ is the 
only parameter controlling the macrostate of a polymer chain.
}

{
We define a forward process as one where an external control $\lambda_\tau$
is varied from $\lambda_0$ to $\lambda_t$ during $0\le\tau\le t$
in a well-defined manner, and a corresponding reverse process $\lambda'_\tau$ as $\lambda'_\tau\equiv\lambda_{t-\tau}$.     
Let $l$ be a space-time trajectory of the system's microstates ({\it i.e.}, evolving polymer configurations) 
during a forward process. Since $l$ fluctuates, we repeat the process $\lambda_\tau$ with appropriate initial probability
density $p(x,0)$ over all microstates $x$. 
When we consider a microstate along a path $l$ at time $\tau$, we will denote it as $l_\tau$.
We define path-dependent work done on the system by the external control during a forward process as follows:
\begin{equation}\label{work}
W (l) \equiv \int_0^t 
\frac{\partial E(l_\tau,\tau)}{\partial \tau} d\tau,
\end{equation}
where $l$ is a trajectory, $E(l_\tau,\tau)$ is the energy of a microstate $l_\tau$ at time $\tau$.
Path-dependent heat transferred to the system during the same process is defined as follows:
\begin{equation}\label{heat}
Q(l) \equiv \int_0^t 
\frac{\partial E(l_\tau,\tau)}{\partial l_\tau} \circ dl_\tau,
\end{equation}
where $\circ \,dl_\tau$ means either $\frac{dl_\tau}{dt}dt$ for deterministic frameworks, or a multiplication
in the Stratonovich sense for stochastic frameworks. Then, for each $l$, by adding the heat and work during the
forward conversion we have the first law of thermodynamics in the following form:
\begin{equation}\label{1st}
\Delta E = W(l) + Q(l), 
\end{equation}
where $\Delta E = E(l_t,t)-E(l_0,0)$.
}

\begin{figure}
\includegraphics[width=7.5cm]{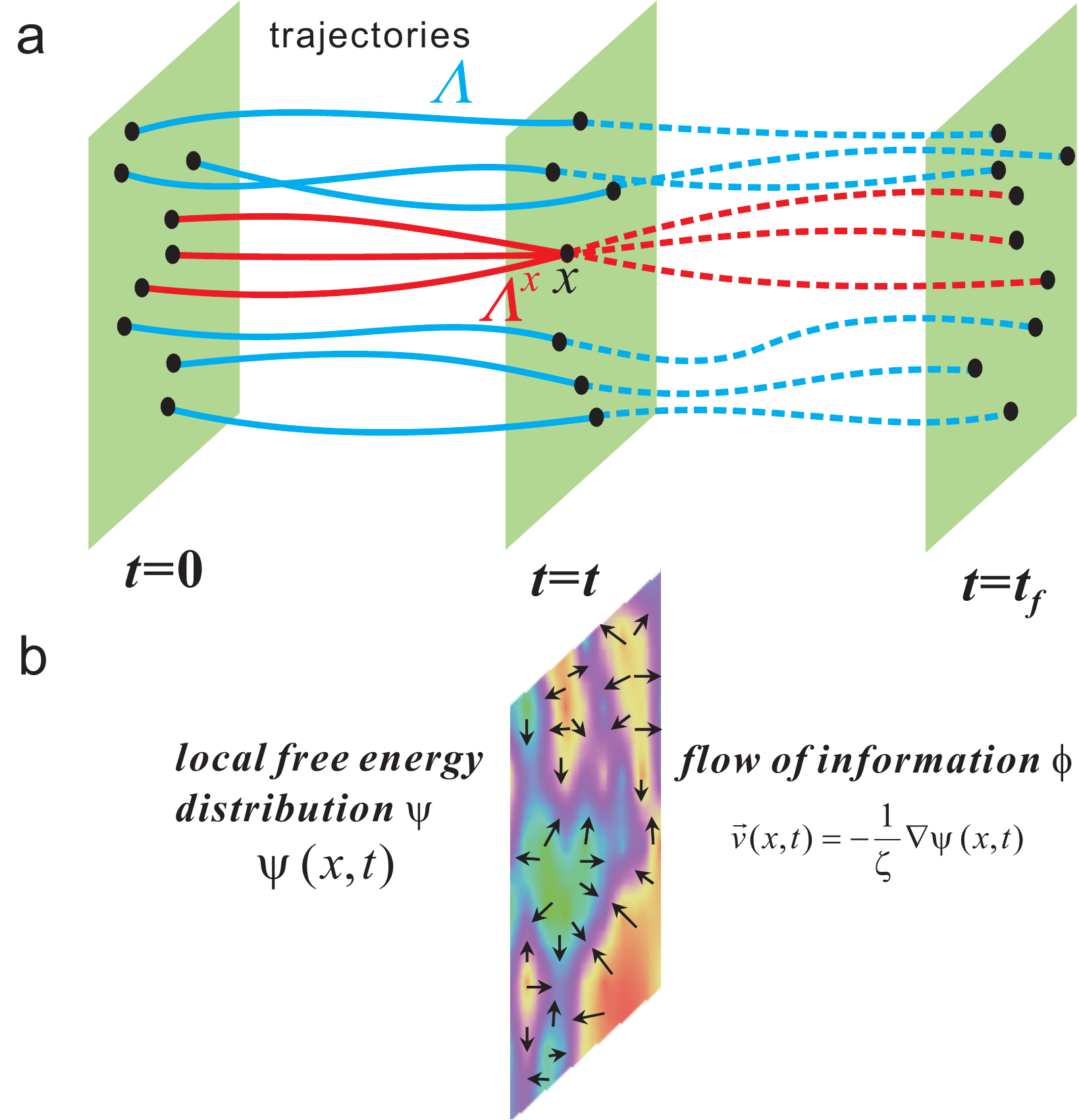}
\caption{
{\bf An evolution of a system's microstates.}
{\bf a,} A microstate as an ensemble of trajectories to a phase-space point. 
Each space in green represents the phase space of a system
(excluding momentum variables for simplicity) at time $t$.
Each trajectory shown schematically represents a stochastic evolution of the phase-space points.   
Among all possible trajectories $\Lambda$, we focus on those paths ($\Lambda^x$: thick red lines)
that reach a specific point $x$ in the phase space at time $t=t$. 
{\bf b,} Schematic figure of the distribution of $\psi(x,t)$ at time $t=t$ in the phase space, where the information flow is to be induced 
in the directions indicated by black arrows. 
The information flow not only causes the convective transport of the information $\phi(x,t)$, but also 
its divergence provides the net change of $\phi(x,t)$ (see Eq. (\ref{eq:dynamics})).} 
 \label{fig:fig2}
\end{figure}

Let $\Lambda$ be the set of all space-time trajectories of a system connecting times 0 and $t$, 
and $\Lambda^{x}$ be those paths that reach a phase space point $x$ at time $t$ 
from the past (see Fig. 2). There is a natural projection $\pi$ from $\Lambda^x$ onto $x$ such that $\pi(\Lambda^x)=x$, and then each microstate
$x$ can be regarded as {an ensemble} 
with conceptual rigour.  {This is because $\pi^{-1}(x)$ includes an {\it infinite} number of stochastic trajectories reaching $x$. 
This indicates that {\it thermodynamic} quantities can be defined for $x$.}  

{
Our results will be built upon two assumptions. 
Firstly, we assume that the probability of path 
$l\in\Lambda$ is represented as $p_\Lambda(l)\propto e^{-s(l)}$  
for some function $s$. The explicit form of $s$ depends on the type of model. For example, it would be the stochastic action with an initial condition derived by Onsager for stochastic models \cite{onsager1}, the Hamiltonian of a system \cite{jarReview}, 
or a bath at time $0$ for deterministic frameworks  \cite{jar2000}. Based on this, we develop the following idea: 
Since each path may not be equally probable, we assign each path a weight of the form 
$g(l)\equiv e^{-s(l)}$ so that less probable paths would be less counted. Let $\Omega(X)$ be the 
{\it accessible number} (or the total weight)  of paths in a sample space $X$, then, we have $\Omega(X) = \sum_{\l\in X} g(l)$.
In this framework, every {\it unit} path, a collection of paths of which total weight
is $1$, is equally probable, and the probability of a microstate ({\it e.g.}, a chain configuration) $x$  at time $t$ 
would be simply the ratio of the accessible number (or the total weight) of trajectories that pass $x$ at time $t$ to 
that of all possible trajectories, {\it i.e.}, $\Omega(\Lambda^x)/\Omega(\Lambda)$. 
This new framework provides us with a basis for ``counting'' analogous to classical entropy.}

{Secondly, we assume microscopic reversibility \cite{crooks99,jar2000}, proven by
Jarzynski in the Hamiltonian framework \cite{jar2000}, and valid in various stochastic frameworks \cite{review}. 
To state this, let us consider a forward process $\lambda_\tau \,(0\le\tau\le t)$, 
and a space-time trajectory $l$ of a polymer chain from a specific configuration (microstate) $l_0$ to another $l_t$ during $t$.
We also consider the time reversed conjugate $l'$ of the path $l$, {\it i.e.,} $l'_\tau\equiv l_{t-\tau}$ 
for $0\le\tau\le t$ under the reverse process $\lambda'_\tau = \lambda_{t-\tau}$.
Here, we set the initial probability density for the reversed process $p'(x,0)$ as the final probability density 
for the forward process $p(x,t)$ so that we have:
\begin{equation}\label{prob}
p'(l'_0,0)=p(l_t,t).
\end{equation}
Now, the condition for microscopic reversibility  \cite{crooks99,jar2000} reads as follows:
\begin{equation}\label{rev}
\frac{p_\Lambda(l \,|\, l_0)}
{p'_\Lambda(l' \,|\, l'_0)} = e^{-\beta Q},
\end{equation}
where $Q$ is heat flowing into the system, 
$\beta=1/k_{\rm B}T$ ($k_{\rm B}$: the Boltzmann constant), 
$p_\Lambda(l\,|\,l_0)$ is the conditional path probability in the sample space $\Lambda$, and $p'_\Lambda(l'\,|\,l'_0)$ 
is that for the reverse process. We note that $p_\Lambda(l\,|\,l_0)\neq p'_\Lambda(l'\,|\,l'_0)$ in general.
}

{Given the above setup, we construct ``local non-equilibrium thermodynamics'', 
where locally-defined entropy, information, and free energy are true thermodynamic potentials
for each microstate, in sharp contrast with ``stochastic thermodynamics'', where they are trajectory-dependent.}

\section*{Results}
\subsection*{Local functions as state functions of $\Lambda^x$}
{
Firstly, we prove that various local functions defined at each
phase-space point $x$ at time $t$ are actually state functions
of the newly introduced ensemble $\Lambda^x$.
Let us consider a local form of information theoretic Shannon entropy (so called stochastic entropy)
\cite{crooks99,seifert05}. To show that this entropy is the property of the ensemble $\Lambda^x$,} we define a quantity for $\Lambda^x$:
\begin{equation}\label{phi}
\phi(x,t)=\ln \Omega(\Lambda^{x}),
\end{equation}
which we shall call {\it information}. As a reference value let $\phi_0$ be the maximally
attainable value of information, {\it i.e.}, ${\phi_0=\ln \Omega(\Lambda)}$ (see Fig. 2a).
Then, ${\sigma(x,t)\equiv k_{\rm B}(\phi_0-\phi(x,t))}$ becomes 
{$ -k_{\rm B}\ln \frac{\Omega(\Lambda^x)}{\Omega(\Lambda)}$. 
Since the argument of the logarithm is simply the ratio of the total weight of paths in $\Lambda^x$ to that of $\Lambda$,
which becomes $p(x,t)$, $\sigma(x,t)$ is identical to the local Shannon entropy. Thus, stochastic entropy $\sigma(x,t)$} 
can be viewed as a property of the ensemble $\Lambda^x$ as desired. 
{Here we emphasize that the local entropy is now defined through a means of counting.} 
We remark that $\phi(x,t)$ 
may be interpreted as the information on the occurrence of the microstate ({\it i.e.}, the polymer configuration), 
{entropy} $\sigma(x,t)$ as {\it lost information} upon specifying a particular microstate $x$. Thus, the sum
of information {$\phi(x,t)$ and entropy $\sigma(x,t)$} conserves {\it locally}. 
Hence, any theory for information could be interpreted using entropy and vice versa, and 
we will select one depending on the context. In this picture, every microstate of a system carries information, and thus we will consider
{information gain/consumption of each microstate but not information measurement/erasure procedures
formulated} in terms of mutual information \cite{lloyd1,lloyd2,sagawa,sagawa2}.
Now we define a free-energy-like quantity as follows: 
\begin{equation}\label{psi}
\psi(x,t)=E(x,t) - T \sigma(x,t).
\end{equation}
{Note that ${\sigma(x,t) = k_{\rm B}(\phi_0-\phi(x,t))}$.
Thus, local free energy $\psi(x,t)$ enables us to treat the information and energy 
of each microstate on an equal footing since $\psi(x,t)$ is nothing but 
the sum of energy $E(x,t)$ and information $\phi(x,t)$ adjusted by the reference value $\phi_0$ and scaled by $k_{\rm B}T$.}

{It was shown by Hummer and Szabo \cite{hummer} that energy $E(x,t)$ could be represented as weighted average
of work functions over all paths in $\Lambda^x$. In detail, they derived the following relation using the Feynmann-Kac formula:
\begin{equation}\label{hummer}
\left< e^{-\beta W}\delta(x,t)\right>_\Lambda = \frac{e^{-\beta E(x,t)}}{e^{-\beta F_{eq}(\lambda_0)}},
\end{equation}
where the bracket indicates an average over all paths in $\Lambda$, $\delta(x,t)$ is the Dirac-delta function
at a microstate $x$ and time $t$, and $F_{eq}(\lambda_0)$ is the equilibrium free energy
of a macrostate $\lambda_0$. Due to the end-point conditioning by $\delta(x,t)$, only the paths in
$\Lambda^x$ among $\Lambda$ contribute to the value of the left-hand side of this relation.
Thus,  \eqreff{hummer} shows that energy $E(x,t)$ is to be a property of the ensemble $\Lambda^x$.  Accordingly, the local free energy 
$\psi(x,t)$ is also a property of $\Lambda^x$ as desired (see Appendix A).
We note that the average of $\psi(x,t)$ over microstates is known as the {\it effective free energy} of a macrostate,
and relations between the effective free energy and 
relative entropy of macrostates have been investigated in \cite{esposito2011}. 
} 

\subsection*{Roles of local free energy $\psi(x,t)$ within $\Lambda^x$}
Now, we derive fundamental relations for each non-equilibrium microstate,
which share their mathematical forms with those of classical and modern theories.

{
Firstly, we show that the newly introduced local free energy $\psi(x,t)$ is
a critical quantity in determining the non-equilibrium probability of each microstate $x$. As above, we
explicitly calculate the ratio of the total weight of paths in $\Lambda^x$ to that in $\Lambda$ 
(see Appendix B).}
Then, from \eqreff{phi} and \eqreff{psi}, we have the following:
\begin{equation}\label{neq2}
p(x,t)=\frac{e^{-\beta E(x,t)}}{e^{-\beta\psi(x,t)}}. 
\end{equation} 
This relation shows that {local free energy} $\psi(x,t)$ plays a role analogous to
equilibrium free energy for non-equilibrium {microstates}. 
{We stress that $p(x,t)$ is expressed solely by (micro-)state functions ({\it i.e.}, true thermodynamic potentials)    
of the ensemble $\Lambda^x$, which is independent of specific paths to realize a microstate $x$}. 
Here it may be worth noting that the structure of our theory is not dependent on the resolution used for specifying a microstate (see Methods). 
In the literature, \eqreff{neq2} appears implicitly either from definitions of path-dependent stochastic entropy and 
free energy, or mixed with work functions (path functions) if an ensemble is considered \cite{crooks2000, jar_lag2009}.
For the latter, $p(x,t)$ could be written in the above simple form as a property of the ensemble $\Lambda^x$ 
only if the relation between $\psi(x,t)$ and work functions were resolved as shown later in \eqreff{workpsi2}.

{
Secondly, we prove a {\it local} version of the Crooks relation.
The original Crooks relation is a core equation that
can generate various modern fluctuation theorems 
within $\Lambda$ focusing on implementing a macrostate 
$\lambda_t$ from an initial ensemble \cite{crooks2000}.
The {\it local} Crooks relation that we prove holds 
while implementing a single microstate, and thus it can generate various fluctuation theorems within $\Lambda^x$ as will be shown below. 
Using \eqreff{rev} and \eqreff{neq2}, we have
\begin{equation}
\frac{p_\Lambda(l)}{p'_\Lambda(l')}=e^{\beta(\Delta E - \Delta\psi - Q)},
\end{equation}
where $\Delta E=E(l_t,t)-E(l_0,0)$, and $\Delta\psi = \psi(l_t,t)-\psi(l_0,0)$.
Applying the first law as in \eqreff{1st}, we have ${\frac{p_\Lambda(l)}{p'_\Lambda(l')}=e^{\beta(W-\Delta\psi)}}$.
Now we consider path probabilities within $\Lambda^x$ so that $p_\Lambda(l)=p(l_t,t) p_{\Lambda^x}(l)$
and $p'_\Lambda(l')=p'(l'_0,0) p'_{\Lambda^x}(l')$. Due to \eqreff{prob}, we obtain the following \emph{local} Crooks relation:
\begin{equation} \label{crooks}
\frac{p_{\Lambda^x}(l)}{p'_{\Lambda^x}(l')}=e^{\beta(W-\Delta\psi)}.
\end{equation}
This relation implies that fluctuation theorems for (realizing) a macrostate remain the same even if we focus on a single microstate. 
Most importantly, the local Crooks relation provides a physical interpretation for local free energy $\psi(x,t)$. 
To see this, we integrate \eqreff{crooks} over the ensemble $\Lambda^x$, giving:
\begin{equation} \label{workpsi}
\left<e^{-\beta(W-\Delta\psi)}\right>_{\Lambda^x}=1,
\end{equation}
where the bracket indicates an average taken over all paths to $x$ at time $t$.
If we assume that $p(x,0)$ is the Boltzmann-Gibbs distribution (then, we have $\psi(x,0)=F_{eq}(\lambda_0)$ for all $x$ 
(see \eqreff{fluct1} below)), we may rewrite \eqreff{workpsi} as follows:
\begin{equation}\label{workpsi2}
{\left<e^{-\beta W}\right>_{\Lambda^x}=e^{-\beta\Delta\psi}}. 
\end{equation}
{This is nothing but the local form of Jarzynski's relation, expressed for each microstate, revealing that the local free 
energy $\psi(x,t)$ encodes (regulates) work contents for realizing a single microstate $x$ from an initial ensemble.
This role of $\psi(x,t)$ is analogous to that of equilibrium free energy $F_{eq}(\lambda_t)$, as expected.}
As a corollary, we have 
\begin{equation}\label{eq:coro}
\left<W\right>_{\Lambda^x}\geq \Delta\psi,
\end{equation}
which shows} that average efficiency of the conversion from work to $\psi$ is less than 100\%,  
indicating that  the second law of thermodynamics holds even within the ensemble of 
realizations of each microstate $x$. 
It should be noted that local ensemble $\Lambda^x$ is critical to reveal
this role of $\psi(x,t)$: {This term would otherwise be integrated out to give equilibrium free energy
(see \eqreff{fluct1}).
We remark that \eqreff{workpsi2} is a highly desired generalization of Landauer's principle, 
which quantifies fluctuations in erasure processes
\cite{memory, berut} (see Examples and also Discussion).
} 

\subsection*{Roles of local free energy $\psi(x,t)$ between $\Lambda^x$s}
Finally, we prove two inter-microstate relations. The first relation explains how non-equilibrium work measurement could
give equilibrium free energy. Since the {\it accessible number} of microstates, $\int e^{-\beta E(x,t)}dx$, is a function 
of an equilibrium free energy $F_{eq}(\lambda_t)$ of a state $\lambda_t$, \eqreff{neq2} implies 
the following link between the instant distribution for local free energy $\psi$ and an equilibrium free energy for the corresponding macrostate:
\begin{eqnarray}\label{fluct1} 
\left<e^{-\beta\psi(x,t)}\right>_x=e^{-\beta F_{eq}(\lambda_t)},
\end{eqnarray}
where the bracket indicates the average over all microstates (chain configurations). 
According to \eqreff{workpsi2}, non-equilibrium work measurement within $\Lambda^x$ gives 
non-equilibrium free energy $\psi(x,t)$. 
Thus if the measurement is done over all paths in $\Lambda$, it corresponds to taking the average of $\psi$,
which gives equilibrium free energy as expressed in \eqreff{fluct1}.
As a corollary, we have ${\left< \psi(x,t)\right>_x \ge F_{eq}(\lambda_t)}$.
In equilibrium, ${\psi(x,t)=F_{eq}(\lambda_t)}$ holds for all $x$ due to the strict convexity of
the exponential function \cite{cover}. 
This condition together with \eqreff{neq2} gives the Boltzmann-Gibbs distribution, and
clearly characterizes the meaning of the ``{\it least biased}'' in an equilibrium {as presented} by Jaynes \cite{jaynes}. 
{Boltzmann assigned each microstate an equilibrium thermal quantity: 
The above equilibrium condition for $\psi(x,t)$ shows that this is valid only in equilibrium.} 

{
Next, we demonstrate the critical role of local free energy $\psi(x,t)$ in non-equilibrium relaxation. 
We consider} a Brownian system described by the following Langevin equation, which 
is a minimal prototype that contains the stochasticity: $\zeta \dot{x}=-\nabla E(x,t) + \xi$, 
where $\zeta$ is the friction constant and $\xi$ is the fluctuating force that satisfies the fluctuation-dissipation theorem, 
$\langle\xi(t)\xi(t')\rangle=2k_{\rm B}T \zeta \delta(t-t')$. Then, the probability density $p(x,t)$ of this system 
 obeys the Fokker-Planck equation as
${\zeta \frac{\partial p(x,t)}{\partial t}=
\frac{\partial\left( {\nabla E(x,t)p(x,t)} \right)}{\partial x}+k_{\rm B}T\frac{\partial^{2}p(x,t)}{\partial x^2}.}$
From \eqreff{neq2}, the time evolution of local information $\phi(x,t)$ is described by 
\begin{equation} \label{eq:dynamics}
\frac{\partial \phi(x,t)}{\partial t}+\vec{v}(x,t) \cdot \nabla \phi(x,t) = - \nabla\cdot \vec{v}(x,t),
\end{equation}
where $\vec{v}(x,t) = -(1/\zeta)\nabla\psi(x,t)$ (see Appendix C). 
We note that this flow driven by local free energy gradient should be interpreted as information flow (not entropy flow) (see Fig. 2b).
When information flows from $x$ to $x'$ with a loss, we may say that entropy flows with an increase to 
the opposite direction, $x'$ to $x$, since the sum of them is conserved locally.
{Thus, \eqreff{eq:dynamics} tells us that for regions of high work content, $\psi$ spontaneously decreases through the exchange of} local
information and entropy until all available sources of energy are consumed {
so that driving force $\psi(x,t)$ becomes constant for all $x$, {\it i.e.}, $E(x,t)=T\sigma(x,t)$ for all $x$ (up to an additive constant).} 
In a non-equilibrium steady state, the relation $\vec{v} \cdot \nabla \phi +\nabla \cdot \vec{v}=0$ should be satisfied.
Here we emphasize that the above equation is a closed form partial differential equation, 
which is for the information content between ensembles $\Lambda^x$, 
{{\it i.e.}, the quantity independent of individual paths, or (\emph{micro-})\emph{state functions}.
This leads to essential differences from apparently similar dynamical equations, {\it e.g.,} the ordinary differential  
evolution equation for \emph{macroscopic} Shannon-entropy \cite{peliti} and the ordinary differential equation for 
entropy production \emph{along stochastic paths} in \cite{seifert05}.}

\subsection*{Examples}
{We present two examples of non-equilibrium systems where the concept of local free energy allows elegant representation 
of all relevant phenomena.} 

{
In the first example, we will see that local free energy $\psi(x,t)$ provides a new insight for understanding
the cyclic behaviour of the famous Szilard engine \cite{leff2002}.
In step (1), a single particle in a box of volume $V$ in contact with a heat bath of 
temperature $T$ is in equilibrium. For analysis, we will split the phase-space of the particle into two:
the left-half of the box and the right-half denoted as $L$ and $R$, respectively, and 
use this coarse-grained phase-space (see Methods).
Let $\beta=1$ for a while for simplicity.
Then, the free energy of the system is $F_{eq}=-\ln Z,$ where $Z = \int_V e^{-E(x,t)} dx$.
In step (2), a partition is inserted in the middle of the box.
Before measurement, the local free energy of the left and
the right would be the same,   
$\psi(L) = \psi(R) = -\ln Z_1 - \ln 2$,
where $Z_1=\int_{V/2} e^{-E(x,t)} dx$ (see \eqreff{eq:cgfree}). 
Note that the local free energy is not different from the initial equilibrium free energy at this point,
{\it i.e.}, $\psi(L) = \psi(R) = F_{eq}$.
In step (3), we measure where the particle is, and assume that it was left without loss of generality.
Then, we would have $\psi(L)=-\ln Z_1$. Since $\psi$ is a true thermodynamic potential,
independent of paths realizing a state $L$ as shown in \eqreff{workpsi2}, we can deduce that work should be 
consumed upon measurement such that $\left<W\right>_L \ge \Delta\psi = \ln 2$ on
average (\eqreff{eq:coro}). In step (4), we link a weight in accordance with the
observation in step (3) and convert heat into work by the amount of $\ln 2$. 
Thus, it does not violate the second law of thermodynamics.
}

\begin{figure*}
\centering
\includegraphics[width=14.5cm]{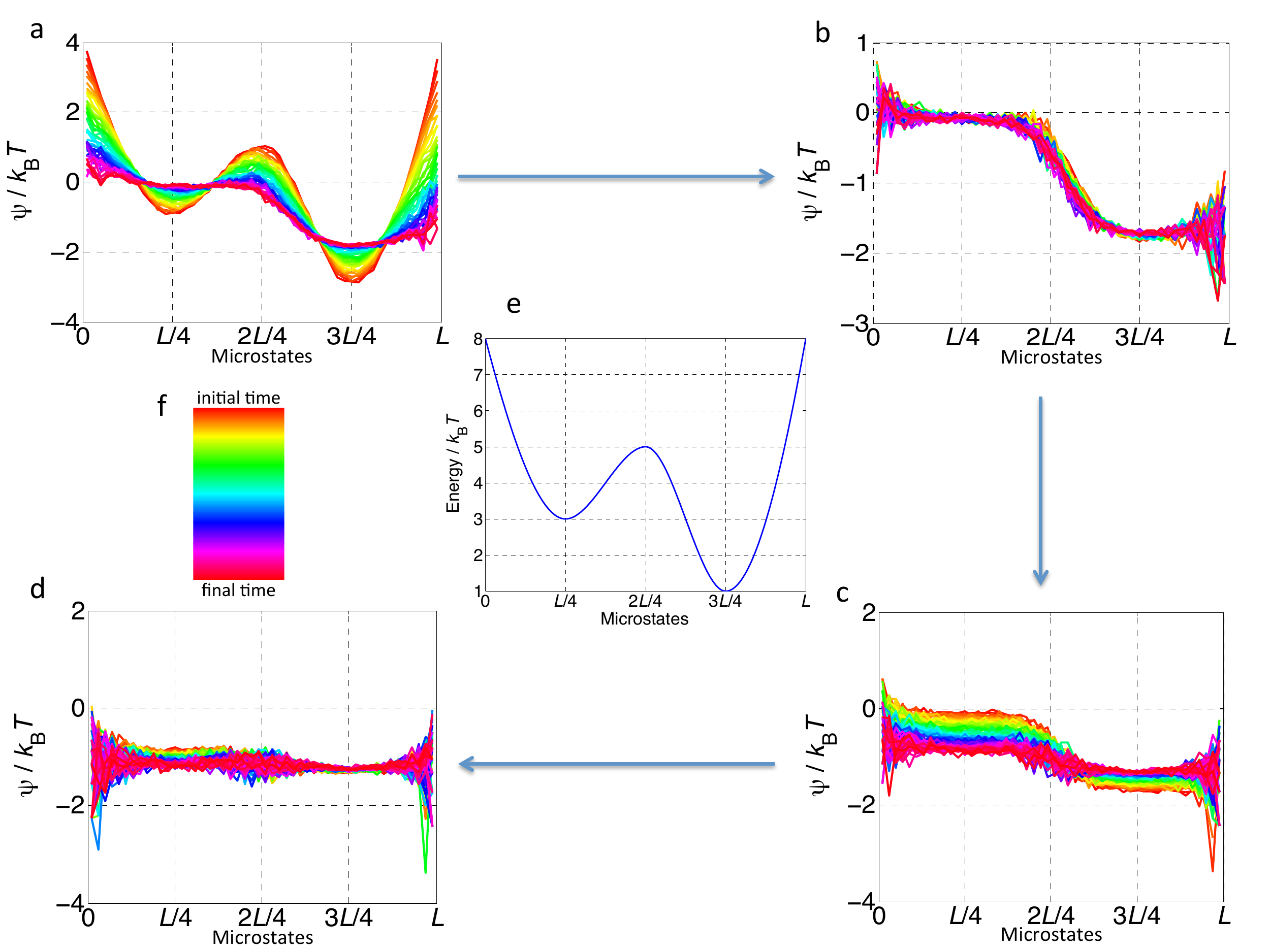}
\caption{{\bf The mechanism of equilibration.}
Here, we describe a stochastic system in terms of local information and 
local free energy extracted from observations of  
stochastic trajectories, and show the exact mechanism of equilibration.
A stochastic evolution in silico of a Brownian particle in a biased potential ({\bf e})  
is carried out 20,000 times with uniform initial distribution, and the time series of local free energy 
$\psi(x,t_i)$ profiles are shown. The colour code is shown in 
{\bf f}. The time-step taken is 0.01 in a dimensionless unit.
{\bf a}-{\bf d} shows the profiles of free energy
$\psi(x,t_i)$ from $t_0$ to $t_{50}$, from $t_{51}$ to $t_{100}$,
from $t_{101}$ to $t_{2000}$, and from $t_{2001}$ to $t_{4000}$, respectively.
{\bf a,} Due to the energy barrier in the transition state 
information flows into local minimum regions. 
As it progresses, the barrier is lowered. {\bf b,} A local equilibrium is established  
where the probability of microstates is proportional to the Boltzmann factor
in local regions. At this stage, the bump has disappeared.
{\bf c,} A global equilibration proceeds slowly compared to  the first local-equilibration process. 
The work potential $\psi(x,t)$ drives information to the right region.
{\bf d,} A global equilibrium is established. Details of the simulation and results with 
different initial conditions are reported in Appendix D.
}
\end{figure*}

{
Here we note that a similar argument also applies to Landauer's erasure process \cite{landauer}. In detail, the first and 
the final stage of the erasure process are identical to step (2) and step (3) of the above example of the Szilard engine, respectively. 
In this case, however, step (3) is achieved not by measurement but by perfectly localizing the particle 
to the left irrespective of its initial position 
(e.g., by extracting the partition separating the two halves of the box and subsequently using a piston to shift the particle to the left side of the box).
In any case, consumed work is characterized by the state function $\psi(L)$, independently of paths to 
realize the state. We remark that fluctuations of work (not just average work)
are exactly regulated by $\Delta\psi$, as expressed in \eqreff{workpsi2}.}

{In the second example,} we illustrate how our theory converts raw data collected from fluctuating degrees of freedom
into a useful set of {state functions as well as the important role of local free energy $\psi(x,t)$
in non-equilibrium relaxation, both} by means of Brownian dynamics simulations (see Appendix D).
We consider a Brownian particle under a tilted bi-stable potential (Fig. 3e), which has a uniform distribution at time $t = 0$.
We repeated the in-silico experiment, counted the number of particles for each partitioned bin at each time, and
calculated local information and free energy.  Recall that information is the accessible number
of paths to a point $x$. Due to the {imposed} initial condition of constant local information, the {local} free energy $\psi$ 
initially has the same shape as the energy $E(x,t)$ over all microstates (Fig. 3a). We observe 
that it is $\psi$ that drives the information flow to establish local equilibrium as an intermediate step (Fig. 3b)
until global equilibrium is reached (Fig. 3d). This is an illustration of an exact mechanism for non-equilibrium
relaxation (as expressed by \eqreff{eq:dynamics}), 
which, {to our knowledge,} has not yet been {achieved} (see the legends of Fig. 3 and Appendix D).

\section*{Discussion}
{
We have identified a new ensemble for each microstate, and revealed the important roles played by local free energy
$\psi(x,t)$ of the new ensemble. Most importantly, $\psi(x,t)$ turns out to be a key quantity for generalizations 
of the Boltzmann-Gibbs distribution and Landauer's principle to include arbitrary fluctuations, 
and controls non-equilibrium relaxation. In the literature, related equations to this manuscript 
appear implicitly in various forms without awareness of the role of the local free energy $\psi(x,t)$.
For example, the most representative form for the generalized Boltzmann-Gibbs distribution appears in
the literature \cite{crooks2000,jar_lag2009} as follows: 
\begin{equation}\label{hummer2}
\frac{p(x,t)}{p_{eq}(x,t)}=\frac{e^{-\beta\Delta F}}{\left< e^{-\beta W}\right>_{\Lambda^x}},
\end{equation} 
where $p_{eq}(x,t)$ is the Boltzmann-Gibbs distribution, $\Delta F = F(\lambda_t)-F(\lambda_0)$ that is the difference
between equilibrium free energy. This relation is a combined form of equations \eqref{neq2},
\eqref{workpsi2}, and a corollary of \eqref{fluct1}. 
In \cite{broeck2014}, \eqreff{neq2} arises simply from the definitions of \emph{path-dependent} entropy and free energy
without a significant relationship between $\psi(x,t)$ and work as given in \eqreff{workpsi2}. 
In \cite{berut}, fluctuations in Landauer's principle have been investigated using \eqreff{hummer2} to find
Landauer's bound without awareness of the role, {\it i.e.}, 
the exact quantification of the fluctuation, 
played by $\psi(x,t)$. 
Very recently, symmetry-breaking energetics ({\it e.g.},  for a Brownian particle 
that is experiencing a switching from a single-well to a double-well
potential and is then confined in one well resulting in loss of ergodicity)   
has been investigated in \cite{roldan2014}, and it has been shown that 
\begin{equation}\label{sym}
\left<W\right>_i^{\rm (SB)} - \Delta F_i \ge k_{\rm B}T\ln p_i,
\end{equation} 
where $\left<W\right>_i^{\rm (SB)}$ is the average work required for symmetry-breaking, 
$\Delta F_i = F_i - F_{eq}(\lambda_0)$, 
where $F_i$ satisfies $e^{-\beta F_i}=\int_i e^{-\beta E(x,t)} dx$ and is 
called conformational free energy of the trapped region $i$ of the phase-space,
and $p_i$ is the probability of trapping in the region $i$.
If we rewrite \eqreff{sym} using the coarse-grained phase-space notation (see Methods)  
and \eqreff{eq:cgprob}, we have
\begin{equation}
\left<W\right>_i^{\rm (SB)} \ge \Delta\psi,
\end{equation}
where $\Delta\psi = \psi(i,t) - F_{eq}(\lambda_0)$. This relation is identical to \eqreff{eq:coro}, 
a corollary of our main theorem, \eqreff{workpsi2}, indicating that local free energy $\psi$ regulates fluctuations in
symmetry-breaking energetics. Overall, these examples indicate the universal importance 
of local free energy $\psi(x,t)$ in the thermodynamic expressions. 
}

{
The second issue we discuss concerns structural similarity between an equilibrium macrostate and a 
non-equilibrium microstate equipped with the new ensemble $\Lambda^x$.
We showed that the Boltzmann-Gibbs distribution and Jarzynski's relation, which hold for (implementing) a macrostate, 
can be expressed in a similar manner while implementing a non-equilibrium microstate.
Moreover, local functions are well-defined state functions of the new ensemble $\Lambda^x$, 
{\it i.e.}, true thermodynamic potentials that are independent of paths of realizing a (micro-)state,
playing similar roles to equilibrium macrostate functions like free energy.
If we pursue this direction further, we may borrow various notions from equilibrium thermodynamics.  
For example, we may define non-equilibrium temperature locally as 
$dE(x,t)/d\sigma(x,t)$. 
In principle, this approach enables us to transfer all the techniques of equilibrium thermodynamics to analyse non-equilibrium local objects. 
We believe that there is significant value in investigating the usefulness of such an approach in non-equilibrium systems.
}

\section*{Methods}
\subsection*{Practical definition of a microstate}
Here we briefly mention a subtlety associated with the definition of a microstate by using a polymer configuration as its example. 
When we try to specify a configuration of a polymer chain at time $t$, there is always a problem of the space-time resolution that is used to distinguish 
different microstates. This is particularly the case for experiments and simulations. 
This resolution problem can be resolved by a coarse-grained partition of the phase space.
Let $\{\chi_j|j=1,\cdots, K\}$ be a partition of the phase space with $K$ non-overlapping subsets $\chi_j$. 
We may regard $\chi_j$ as a coarse-grained microstate. The information of $\chi_j$ is defined as follows:
\begin{equation}
\phi(\chi_j,t)=\ln \Omega(\Lambda^{\chi_j}),
\end{equation}
where $\Omega(\Lambda^{\chi_j})=\sum_{x\in\chi_j} \Omega(\Lambda^x).$ 
Then, we apply the previous argument and obtain the 
non-equilibrium probability of $\chi_j$, $p(\chi_j,t)=e^{\phi(\chi_j,t)-\phi_0}$. 
In this case, we shall define the free energy of $\chi_j$ as follows:
\begin{equation}\label{eq:cgfree}
\psi(\chi_j,t)=E(\chi_j,t) - T\sigma(\chi_j,t),
\end{equation}
where $\sigma(\chi_j,t)=k_{\rm B}(\phi_0-\phi(\chi_j,t))$, {identical to stochastic entropy,} 
and $E(\chi_j,t)$ is defined such that $e^{-\beta E(\chi_j,t)}=\int_{x \in \chi_j} e^{-\beta E(x,t)} dx$.
Then, we have 
\begin{equation}\label{eq:cgprob}
p(\chi_j,t)=\frac{e^{-\beta E(\chi_j,t)}}{e^{-\beta \psi(\chi_j,t)}}.
\end{equation}
{We note two things here: Firstly,} the structure of our theory is invariant under the transformation of a microstate $x$ to
a coarse-grained one $\chi_j$. {Secondly, when we do not divide} fluctuating degrees of freedom, our theory reduces to
global theorems in a natural way. For example, let a single coarse-grained microstate $\chi$ be the set of all fluctuating degrees of freedom.
Then $\psi(\chi,t)$ becomes $F_{eq}(\lambda_t)$ from equation (\ref{eq:cgfree}).
We see immediately that equations {(11)-(14)} in the main text reduce to the corresponding known
global relations, and equation{s (9) and (15) reduce to trivial identities.}
Similarly, it may be instructive to apply the same framework to an equilibrium situation, where a system is in a stationary state 
with thermal fluctuations. Even in an equilibrium situation, we can think of an initial state and a state at time $t$. 
Thus we can still define $\phi(x,t)$ and $\psi(x,t)$. Our intuition immediately tells us that $\sigma(x,t)$ and 
$\psi(x,t)$ {should both be independent of $x$ and $t$, suggesting that their averages are} 
standard entropy and free energy, respectively.

\subsection*{Experimental measurement}
As an example, we consider a polymer chain (in a heat bath of temperature $T$)  
whose one end is fixed and the other end is under an external control $\lambda_{\tau}$ that varies 
over time $0\le\tau\le t$ in a well-defined manner, as shown in the right panel of Fig. 1. 
We prepare the system initially to be sampled  according to a probability density $p(x,0)$.
Let us assume that $p(x,0)$ is the Boltzmann-Gibbs distribution for simplicity.
We carry out the experiment by controlling $\lambda_{\tau}$ during $0\le\tau\le t$, and
measure work $W$ done on the system. If we have additional information on
the final microstate, our theory converts the fluctuating details into $\phi$ and $\psi$
({a microstate can be a coarse-grained one}). We repeat the process $N$ times, {a large number}.
Let $N_x$ be the number of the experiments whose final microstate is $x$. 
Here we put a number to path $i$ reaching microstate $x$ from $i=1$ to $N_x$ 
and express the work done in path $i$ as $W_i$. Then, we obtain the energy of $x$
using the following equality for $\Lambda^x$: 
\begin{equation*}
\frac{1}{N}\sum_{i=1}^{N_x} e^{-\beta W_i} = \frac{e^{-\beta E(x,t)}}{e^{-\beta F_{eq}(\lambda_0)}},
\end{equation*}
{as proven} in \cite{hummer}. Here $F_{eq}(\lambda_0)$ is the equilibrium free energy of
the initial {\it macrostate}.
It is important to note that the normalization factor is $N$. If the summation is taken over all experiments,
it reduces to the Jarzynski relation. We also proved that
\begin{equation*}
\frac{1}{N_x}\sum_{i=1}^{N_x} e^{-\beta W_i} = \frac{e^{-\beta \psi(x,t)}}{e^{-\beta F_{eq}(\lambda_0)}},
\end{equation*}
{giving} $\psi(x,t)$. 
It is important that the normalization factor here is $N_x$.  These two equalities are linked by the 
the relation between $\psi(x,t)$ and $F_{eq}(\lambda_t)$ in \eqreff{fluct1}.

\subsection*{Acknowledgments}
It is a pleasure to thank C. Hyeon and H. Park in KIAS for discussions.
We are also grateful to Taiki Yanagishima for critical reading of our manuscript. 
L.J. was supported by the National Research Foundation of Korea Grant funded by the Korean Government (NRF-2011-013-C00008),
and in part by Kwangwoon University Research Year Grant in 2011. 
H.T. acknowledges Grants-in-Aid for Scientific Research (S) and Specially Promoted Research
from the Japan Society for the Promotion of Science (JSPS), and the Aihara Project, the
FIRST program from JSPS, initiated by the Council for Science and Technology Policy
(CSTP).

\bibliography{bib.bib}

\appendix

\section*{Appendix}
\noindent
{\bf A. Illustration of local $\psi$.}

\begin{figure*}
 \centering
 \includegraphics[width=15cm]{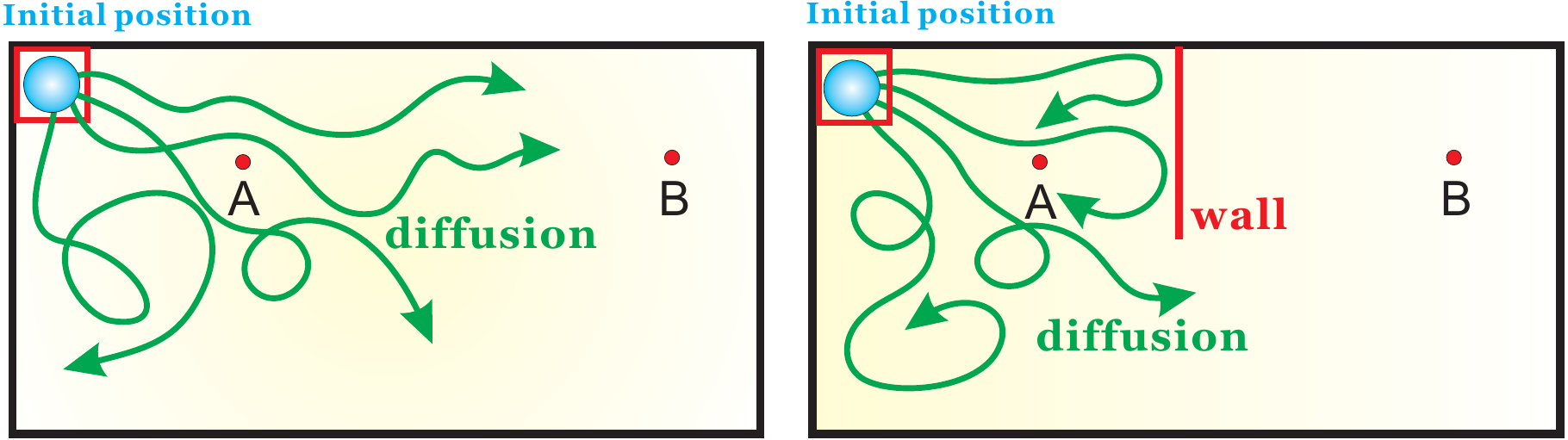}
\caption{{\bf An incompletely-blocked particle. }
A colloidal particle is released at $t=-\epsilon$
for a small $\epsilon>0$
from the upper-left corner so that it has non-vanishing
probability at $t=0$, and moves stochastically. 
We repeat the experiment, and consider to block the particle by a partition incompletely as shown in the right panel. 
During non-equilibrium evolution,
the presence of the wall increases 
the information
content of the ensemble $\Lambda^A$ of A
resulting in an increase of the
local free energy $\psi(A,t)$ for some $t$.
Accordingly, $\psi(B,t)$ would be decreased.
}
\label{fig:psi}
\vspace{1cm}
\end{figure*}

Here, we consider a Brownian particle in a box in contact with a heat bath of temperature $T$
and block it incompletely as shown in Fig. \ref{fig:psi}. 
Let $\Lambda^A$ be all the paths from an initial ensemble
to $A$ at time $t$. We define {\it the accessible number}
(or the total weight) $\Omega(\Lambda^A)$ of paths in 
$\Lambda^A$ by ${\Omega(\Lambda^A)=\sum_{l\in\Lambda^A} g(l)}$, where
$g(l)$ is a weight of a path $l$ such that $g(l)$ is 
proportional to the probability of path $l$ (see the main text, or Section B below).
The information content $\phi(A,t)$ is defined by 
$\phi(A,t)=\ln \Omega(A,t)$. As a reference value $\phi_0$
is defined by the logarithm of the total weight of all paths,
{\it i.e.}, $\phi_0=\ln \Omega(\Lambda)$, where $\Lambda$
is the set of all paths. Now, the presence of the partial wall increases
the accessible number (or the total weight) $\Omega(\Lambda^A)$
of paths to $A$, so does information content $\phi(A,t)$,
and thus the local free energy $\psi(A,t)$. Note that the local free energy is defined
by $\psi(A,t) = E(A,t) - T\sigma(A,t)$, where $E(A,t)$ is the energy of the microstate $A$ at time $t$ and 
$\sigma(A,t) = k_{\rm B}(\phi_0 - \phi(A,t))$.  
Similarly, $\psi(B,t)$ would be decreased. As we will see from the main text, $\psi$ is the work content.
Thus, we could extract work from the increase of $\psi(A,t)$ by linking a weight to the partition
during non-equilibrium evolution of the particle.
We also note that the thermodynamical asymmetry between A and B is due to the initial condition, indicating the importance of 
initial conditions in non-equilibrium \cite{jarReview}. 

\vspace{3mm}
\noindent
{\bf B. Derivation of Eq. (9).}
 
\begin{figure*}
 \centering
 \includegraphics[width=10cm]{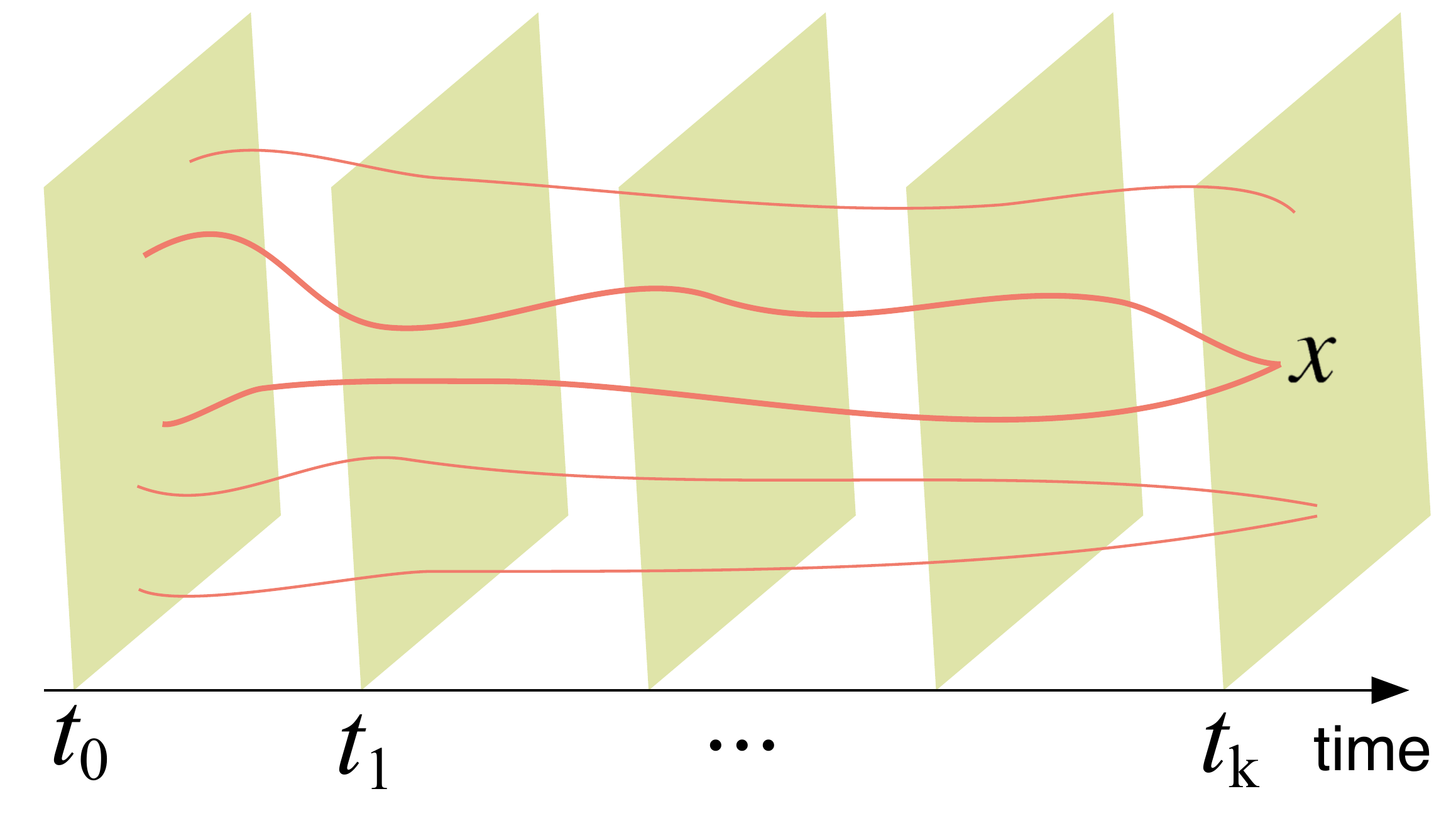}
 \caption{
{\bf A microstate as an ensemble of trajectories to
the phase-space point.} 
Each space in green represents the phase space of a system
(excluding momentum variables for simplicity) at time $t_i$.
Each trajectory shown schematically represents a stochastic evolution of the phase-space points.   
Among all possible trajectories, we focus on those paths (thick lines)
that reach a specific point $x$ in the phase space at time $t_{\rm k}$.}
 \label{fig:path}
\end{figure*}

Let $\Lambda$ be the set of all possible space-time 
trajectories of an experiment.
When we count {\it the accessible number}
of paths, 
we assign each path a weight of the form 
$g(l)\equiv e^{-s(l)}$, where $s(l)$ is such that
the probability of path $l$ is represented
as $p(l)\propto e^{-s(l)}$, so that less probable paths are to be less counted.
Then,  
a probability of path $l$ in a sample space $X$
would be $p(l) = g(l)/\sum_{l\in X} g(l)$.
It would be convenient to deal with a trajectory
in a discrete approximation 
as a set of states $x_i$ in consecutive times $t_i$
for ${i=0,\cdots, n}$.
Let us denote the set of trajectories that pass
$x_k$ at time $t_k$ as $\Lambda(x_k,t_k)$ for some 
integer $k$ (see Fig. \ref{fig:path}).
We will calculate the ratio of the number of paths in
$\Lambda(x_k,t_k)$ to 
the number of all possible trajectories in $\Lambda$.
Since the probability of path $l$ is
$p(l)=g(l)/\sum_{l\in\Lambda} g(l)$,
the ratio becomes $\sum_{l\in\Lambda(x_k,t_k)} p(l)$.
Then, the ratio would be approximated as 
${\sum_{x_0}\cdots\sum'_{x_k}\cdots 
\sum_{x_{n}} p(x_0,\cdots,x_k,\cdots,x_n)}$, where each sum is taken over the phase-space points 
at time $t_i$, and $\sum'_{x_k}$ means that we omit 
the sum over the phase-space points at time $t_k$. 
It is important to omit that sum at time $t_k$,
otherwise the result would be $1$.
By the law of total probability the sum reduces to 
$p(x_k,t_k)$.

\vspace{3mm}
\noindent
{\bf C. Derivation of dynamic rules.} 

We consider the Langevin equation, which 
is a minimal prototype 
that contains an essence of stochasticity: 
$\zeta \dot{x}=-\nabla E(x,t) + \xi$, 
where $E(x,t)$ is energy of a microstate $x$ at time $t$,
$\zeta$ is the friction constant and $\xi$ is the fluctuating force that satisfies the fluctuation-dissipation theorem, 
$\langle\xi(t)\xi(t')\rangle=2k_{\rm B}T \zeta \delta(t-t')$. 
The probability density $p(x,t)$ of a system 
under the overdamped Langevin equation obeys the Fokker-Planck equation as
\begin{equation}\nonumber
\zeta \frac{\partial p(x,t)}{\partial t}=
\frac{\partial\left( {\nabla E(x,t)p(x,t)} \right)}{\partial x}+k_{\rm B}T\frac{\partial^{2}p(x,t)}{\partial x^2}.
\end{equation}
From $p(x,t)=e^{\phi(x,t)-\phi_0}$ 
(see the main text or Sections A and B above), 
we have $\frac{\partial p}{\partial x} = 
p \frac{\partial \phi}{\partial x} $ and
${\frac{\partial^2 p}{\partial x^2}  = p\left(\frac{\partial  
\phi}{\partial x} \right)^2 + p \frac{\partial^2 \phi}
{\partial x^2}}$.
Substituting these relations into the Fokker-Planck equation, and dividing by $p(x,t)$, 
we obtain the following equation: 
\begin{eqnarray}\nonumber
\zeta \frac{\partial \phi(x,t)}{\partial t}
&=&\frac{\partial E(x,t)}{\partial x}
\frac{\partial \phi(x,t)}{\partial x}
+k_{\rm B}T\left(\frac{\partial \phi(x,t)}{\partial x}\right)^2 \nonumber \\
&+& \frac{\partial^2 E(x,t)}{\partial x^2}+
k_{\rm B}T\frac{\partial^2 \phi(x,t)}{\partial x^2}.
\nonumber
\end{eqnarray}
Here the first two terms in the right-hand side
may be written as 
$\frac{\partial}{\partial x}\left(E(x,t) + k_{\rm B}T\phi(x,t)\right) \frac{\partial \phi(x,t)}{\partial x}$,
and the second two terms as $\frac{\partial^2}
{\partial x^2}\left(E(x,t)+k_{\rm B}T\phi(x,t)\right)$.
Noting that the sum in the parenthesis is just $\psi(x,t)$
(see the main text or Section A above), 
we have
\begin{equation}\nonumber
\frac{\partial \phi(x,t)}{\partial t}-\frac{1}{\zeta}\nabla\psi(x,t) \cdot 
\nabla \phi(x,t) = \frac{1}{\zeta} \nabla^2 \psi(x,t),
\end{equation}
where $\nabla^2$ denotes the Laplacian.
This is a non-linear 
convection-diffusion equation with a source term. 
It determines the dynamics
of $\phi$ and $\psi$ completely, given initial and boundary
conditions, where the energetic cost of the information flow is mediated by heat 
with less than 100\% efficiency resulting in a net loss of information, andthus a net increase of entropy.

\vspace{3mm}
\noindent
{\bf D. Details of in-silico experiments.}

We carried out in-silico experiments of the Brownian movement of a particle.  
We numerically solve the following overdamped Langevin equation:
\begin{equation}
\zeta \dot{x}=-\nabla E(x,t) + \xi, \nonumber
\end{equation}
where $E(x,t)$ is energy of a microstate $x$, and thermal fluctuation $\xi$ satisfies $\langle\xi(t)\xi(t')\rangle=2k_{\rm B}T \zeta\delta(t-t')$. 
Here we set $\zeta=1$ for simplicity. Then, after discretization, we have 
\begin{equation}
x(t_{i+1}) = x(t_i) - \nabla E(x(t_i),t_i)  \epsilon + \sqrt{2k_{\rm B}T\epsilon}r(t_i), \nonumber 
\end{equation}
where $\epsilon = t_{i+1}-t_i$ and $r(t_i)$ is a random number drawn from the standard normal distribution.
In our simulations, we set the mobility to unity, and $k_{\rm B}T$ to $2$ by rescaling. 
We constrained the particle by setting the reflecting boundaries at $x=0$ and $x=L$ ($L=10$).
We used $\epsilon=0.01$. 
The domain is partitioned into $50$ bins, and we counted the number of particles for each bin at each time
to obtain the graphs of information and free energy.

\begin{figure*}
\includegraphics[width=6in]{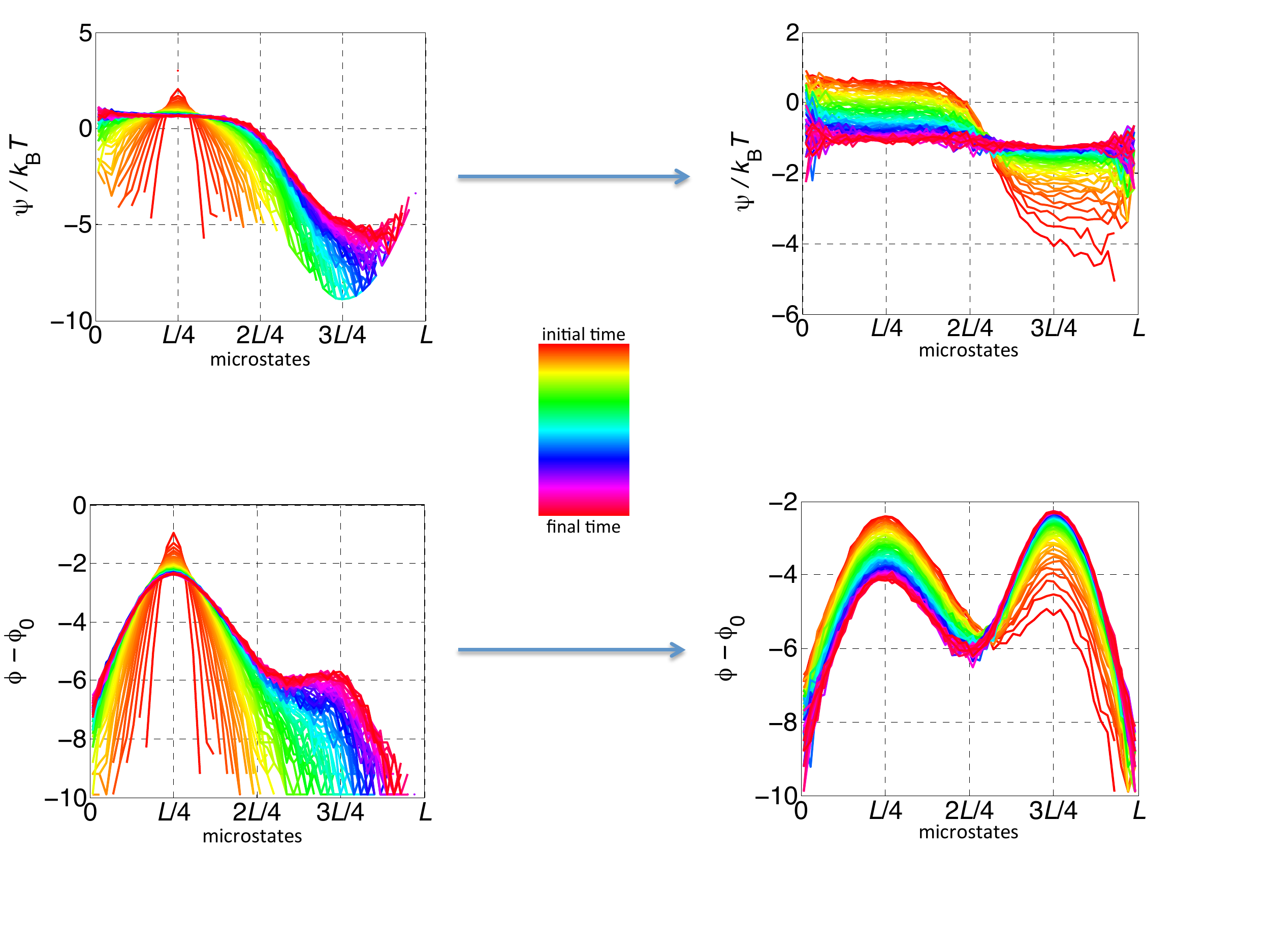}
\caption{{\bf The mechanism of equilibration: case I. }
Here we put the particle only at $x=L/4$ during the repetition of the simulation so that
$p(x,0)=\delta(x-L/4)$. The other conditions are the same as the case in the main text. 
The time series of free energy over microstates
is shown in the upper panel, and those of information
in the lower panel.
The profiles are from $t_0$ to $t_{100}$ for the left figures, and $t_{101}$ to $t_{4000}$ for the right figures.
Due to the initial condition,
there is no free energy barrier in this case. The process towards local equilibrium is very quick.
The global equilibrium proceeds without breaking the established local equilibrium.
}
\label{figs1}
\end{figure*}

Figure \ref{figs1} shows the profiles of information and free energy 
when the initial condition is set to $p(x,0)=\delta(x-L/4)$.
In this case, there is no barrier in $\psi(x,t)$
although there is an energy barrier. 
Thus, the local equilibrium is established quickly in the initial stage. 
Then the free energy $\psi$ drives information $\phi$ to the right 
region until reaching the global equilibrium. 
During this second stage, the flow continues 
without breaking the established local equilibrium.
Note that the profile of information evolves from the hat shape to the shape 
that exactly compensates the energy profile up to an additive constant. 

\begin{figure*}
\includegraphics[width=6in]{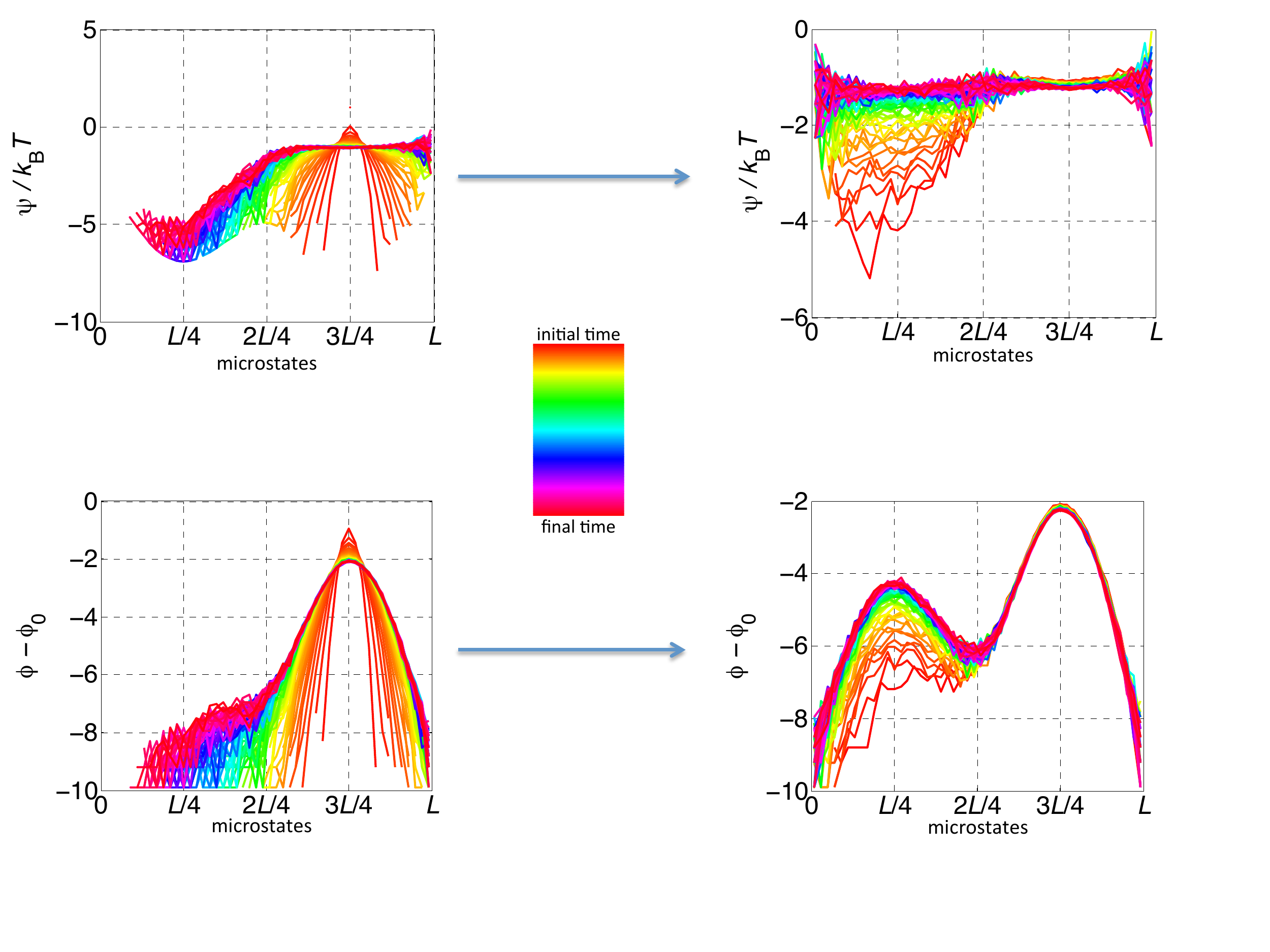}
\caption{{\bf The mechanism of equilibration: case II. }
Here we put the particle only at $x=3L/4$ during the repetition of the simulation so that
$p(x,0)=\delta(x-3L/4)$. The other conditions are the same as the case in the main text.
The time series of free energy over microstates 
is shown in the upper panel, and those of information
in the lower panel.
The profiles are from $t_0$ to $t_{100}$ for the left figures, and $t_{101}$ to $t_{4000}$ for the right figures.
Due to the initial condition,
there is no free energy barrier in this case. The process towards local equilibrium is very quick.
The global equilibrium proceeds without breaking the established local equilibrium.}
\label{figs2}
\end{figure*}

Now we put initially the particle at the location of the global minimum of energy, {\it i.e.} $p(x,0)=\delta(x-3L/4)$,
and Fig. \ref{figs2} shows the profiles of information and free energy over microstates. 
There is no barrier in the free energy profile $\psi$.
First, the local equilibrium is established quickly, 
and the flow of information $\phi$ continues towards the left region
although the speed is much slower than the local 
equilibration process. 
Second, the flow of information $\phi$ continues until
the global equilibrium is established without breaking the local equilibrium.
In this case again, the profile of information 
at the equilibrium compensates exactly the 
energy profile up to an 
additive constant.

\end{document}